\documentclass[twocolumn,times]{aastex62}
\usepackage{natbib}

\newcommand{\kms}{km\,s$^{-1}$}
\newcommand{\cii}{[\ion{C}{2}]}
\newcommand{\mgii}{\ion{Mg}{2}}
\newcommand{\lfir}{$L_{\mathrm{FIR}}$}
\newcommand{\ltir}{$L_{\mathrm{TIR}}$}
\newcommand{\lcii}{$L_\mathrm{[CII]}$}
\newcommand{\lbol}{$L_\mathrm{bol}$}
\newcommand{\lsun}{$L_\sun$}
\newcommand{\msun}{$M_\sun$}

\newcommand{\msunyr}{$M_\sun$\,yr$^{-1}$}
\newcommand{\mdust}{$M_{\mathrm{dust}}$}
\newcommand{\tdust}{$T_{\mathrm{dust}}$}

\received{2018 June 15}
\revised{2018 August 27}
\accepted{2018 September 2}
\shorttitle{Dust emission in an accretion-rate-limited sample of $z\gtrsim6$ quasars}
\shortauthors{Venemans et al.}

\begin{document}

\title{{\Large Dust emission in an accretion-rate-limited sample of $z\gtrsim6$ quasars}}

\correspondingauthor{Bram P.\ Venemans}
\email{venemans@mpia.de}

\author{Bram P.\ Venemans}
\affiliation{Max-Planck Institute for Astronomy, K{\"o}nigstuhl 17, D-69117 Heidelberg, Germany}

\author{Roberto Decarli}
\affiliation{Osservatorio di Astrofisica e Scienza dello Spazio di Bologna, via Gobetti 93/3, I-40129 Bologna, Italy}
\affiliation{Max-Planck Institute for Astronomy, K{\"o}nigstuhl 17, D-69117 Heidelberg, Germany}

\author{Fabian Walter}
\affiliation{Max-Planck Institute for Astronomy, K{\"o}nigstuhl 17, D-69117 Heidelberg, Germany}
\affiliation{Astronomy Department, California Institute of Technology, MC105-24, Pasadena, CA 91125, USA}
\affiliation{National Radio Astronomy Observatory, Pete V. Domenici Array Science Center, P.O. Box 0, Socorro, NM 87801, USA}

\author{Eduardo Ba{\~n}ados}
\affiliation{The Observatories of the Carnegie Institution for Science, 813 Santa Barbara Street, Pasadena, CA 91101, USA}

\author{Frank Bertoldi}
\affiliation{Argelander Institute for Astronomy, University of Bonn, Auf dem H\"ugel 71, D-53121 Bonn, Germany}

\author{Xiaohui Fan}
\affiliation{Steward Observatory, The University of Arizona, 933 North Cherry Avenue, Tucson, AZ 85721-0065, USA}

\author{Emanuele Paolo Farina}
\affiliation{Department of Physics, Broida Hall, University of California, Santa Barbara, CA 93106-9530, USA}

\author{Chiara Mazzucchelli}
\affiliation{Max-Planck Institute for Astronomy, K{\"o}nigstuhl 17, D-69117 Heidelberg, Germany}

\author{Dominik Riechers}
\affiliation{Cornell University, 220 Space Sciences Building, Ithaca, NY 14853, USA}

\author{Hans-Walter Rix}
\affiliation{Max-Planck Institute for Astronomy, K{\"o}nigstuhl 17, D-69117 Heidelberg, Germany}

\author{Ran Wang}
\affiliation{Kavli Institute of Astronomy and Astrophysics at Peking University, 5 Yiheyuan Road, Haidian District, Beijing 100871, China}

\author{Yujin Yang}
\affiliation{Korea Astronomy and Space Science Institute, Daedeokdae-ro 776, Yuseong-gu Daejeon 34055, South Korea}

\begin{abstract}
We present Atacama Large Millimeter Array 1\,mm observations of the rest-frame far-infrared (FIR) dust continuum in 27 quasars at redshifts $6.0\lesssim z<6.7$. We detect FIR emission at $\gtrsim3\sigma$ in all quasar host galaxies with flux densities at $\sim$1900\,GHz in the rest-frame of $0.12<S_\mathrm{rest,\,1900\,GHz}<5.9$\,mJy, with a median (mean) flux density of 0.88\,mJy (1.59\,mJy). The implied FIR luminosities range from \lfir\,$=(0.27-13)\times10^{12}$\,\lsun, with 74\% of our quasar hosts having \lfir\,$>10^{12}$\,\lsun. The estimated dust masses are \mdust\,$=10^7-10^9$\,\msun. If the dust is heated only by star formation, then the star formation rates in the quasar host galaxies are between 50 and 2700\,\msunyr. In the framework of the host galaxy--black hole coevolution model a correlation between ongoing black hole growth and star formation in the quasar host galaxy would be expected. However, combined with results from the literature to create a luminosity-limited quasar sample, we do not find a strong correlation between quasar UV luminosity (a proxy for ongoing black hole growth) and FIR luminosity (star formation in the host galaxy). The absence of such a correlation in our data does not necessarily rule out the coevolution model, and could be due to a variety of effects (including different timescales for black hole accretion and FIR emission).
\end{abstract}


\section{Introduction} 
\label{sec:introduction}

Luminous quasars are powered by accretion onto supermassive black holes with mass $\gtrsim$$10^{8-9}$\,\msun. Such luminous quasars have been found out to very high redshift \citep[the current quasar record holder is at $z=7.54$ at a cosmic age of 690\,Myr after the Big Bang;][]{ban18a}. In the local universe relations have been found between the mass of the central black hole and both the mass of the bulge \citep[e.g.,][]{kor13} and the mass of the galaxy \citep[e.g,][]{rei15}. If such relations were already in place at high redshift, the host galaxies of the distant quasars would be among the most massive galaxies at these early epochs. 

Due to the bright central source, detecting the host galaxy of luminous distant quasars in the rest-frame UV or optical has proven to be very challenging \citep[e.g.,][]{dec12,mec12}. On the other hand, already more than a decade ago studies at rest-frame far-infrared (FIR) wavelengths (redshifted to the observed (sub-)mm) have revealed intense FIR emission coming from the quasar host galaxies \citep[e.g.,][]{pri01,ber03a,wal03,mai05,bee06}. Studying the quasar host galaxies in the (sub-)mm would therefore allow one to characterize the build-up and formation models of massive galaxies. Early bolometer work 
showed that $\sim$30\% of the $z\gtrsim6$ quasars from the Sloan Digital Sky Survey (SDSS) were individually detected with flux densities of $S_\mathrm{obs,\,250\,GHz}\gtrsim1.2$\,mJy \citep{wan08b}, indicating ultraluminous infrared galaxy (ULIRG)-like FIR luminosities of \lfir\,$>10^{12-13}$\,\lsun. Consequently, early efforts to characterize the host galaxies of $z\sim6$ quasars concentrated on these FIR luminous quasars \citep[e.g.,][]{ber03b,wal03,wal04,wal09b,rie09,wan10,wan11a,wan13}. 

Later, \citet{omo13} and \citet{wil13,wil15} followed up a sample of quasars from the Canada-France-Hawaii Quasar Survey (CFHQS) with lower luminosities ($\sim$2\,mag fainter) than the SDSS quasar. They found that the lower-luminosity quasars are, on average, fainter than the SDSS quasars with $S_\mathrm{obs,\,250\,GHz}<1$\,mJy. At the same time, our group started a pilot project targeting all quasars at $z>6.5$ in the (sub-)mm. Initial results showed that luminous quasars can have a range of properties \citep{ven12,ven17a,ban15b} compared to the well-studied SDSS quasars. In \citet{ven16} we reported a tentative correlation between FIR luminosity (a proxy for star-formation) and the bolometric luminosity (a proxy for black hole growths) of the quasars. 

To obtain a less biased view of the host galaxies of $z>6$ quasars and to investigate how the quasar bolometric and FIR luminosities relate to each other, we targeted a UV luminosity-limited quasar sample with the Atacama Large Millimeter Array (ALMA). The properties of the \cii\,158\,$\mu$m emission lines detected from the quasars in our sample are published in \citet{dec18}. From the quasar sample, 85\% were detected in \cii\ at a significance of $>$5$\sigma$, with typical luminosities of \lcii\,$=10^{9-10}$\,\lsun\ \citep{dec18}. In \citet{dec17} we reported the discovery of \cii-emitting companion sources near some of our quasars. In this paper, we will focus on the dust continuum emission of our sample of high-redshift quasars. The paper is organized as follows. In Section~\ref{sec:sample} we will introduce the sample followed by a brief description of the ALMA observations in Section~\ref{sec:observations} and the literature sample in Section~\ref{sec:literature}. Our results are described in Section~\ref{sec:results} and the derived characteristics are discussed in Section~\ref{sec:discussion}. Finally, we summarize our findings in Section~\ref{sec:summary}. 

Throughout this paper all magnitudes are on the AB system. We adopt a concordance cosmology with $\Omega_M=0.3$, $\Omega_\Lambda=0.7$, and $H_0=70$\,\kms\,Mpc$^{-1}$, which is consistent with measurements from {\it Planck} \citep[][]{pla16}. Star-formation rates (SFRs) are calculated assuming a \citet{kro03} initial mass function.

\section{The Sample and New Observations}

\subsection{A Quasar Luminosity-limited Sample}
\label{sec:sample}

\begin{deluxetable*}{lcccccc}[t]
\tablecaption{Properties of the Quasars Targeted in the ALMA Survey. \label{tab:sample}}
\tablewidth{0pt}
\tablehead{ \colhead{Name} & \colhead{R.A.\ (J2000)} &
\colhead{Decl.\ (J2000)} & \colhead{Redshift} &
\colhead{Method\tablenotemark{a}} & \colhead{$M_{1450}$} &
\colhead{References\tablenotemark{b}} }
\startdata
P007+04     & 00$^\mathrm{h}$28$^\mathrm{m}$06$.\!\!^\mathrm{s}$560 &  +04$^\circ$57$^\prime$25\farcs39 & 6.0008$\pm$0.0004 &  \cii\   & --26.58 & 1, 2 \\
P009--10    & 00$^\mathrm{h}$38$^\mathrm{m}$56$.\!\!^\mathrm{s}$527 & --10$^\circ$25$^\prime$54\farcs08 & 6.0039$\pm$0.0004 &  \cii\   & --26.50 & 1, 2 \\
J0046--2837 & 00$^\mathrm{h}$46$^\mathrm{m}$23$.\!\!^\mathrm{s}$662  & --28$^\circ$37$^\prime$47\farcs44 &   5.99$\pm$0.05   & Template & --25.42 & 3, 3 \\
J0142--3327 & 01$^\mathrm{h}$42$^\mathrm{m}$43$.\!\!^\mathrm{s}$710  & --33$^\circ$27$^\prime$45\farcs55 & 6.3379$\pm$0.0004 &  \cii\   & --27.76 & 1, 2 \\
P065--26    & 04$^\mathrm{h}$21$^\mathrm{m}$38$.\!\!^\mathrm{s}$048 & --26$^\circ$57$^\prime$15\farcs60 & 6.1877$\pm$0.0005 &  \cii\   & --27.21 & 1, 2 \\
P065--19    & 04$^\mathrm{h}$22$^\mathrm{m}$00$.\!\!^\mathrm{s}$999 & --19$^\circ$27$^\prime$28\farcs63 & 6.1247$\pm$0.0006 &  \cii\   & --26.57 & 1, 2 \\
J0454--4448 & 04$^\mathrm{h}$54$^\mathrm{m}$01$.\!\!^\mathrm{s}$789  & --44$^\circ$48$^\prime$31\farcs26 & 6.0581$\pm$0.0006 &  \cii\   & --26.41 & 1, 2 \\
J0842+1218  & 08$^\mathrm{h}$42$^\mathrm{m}$29$.\!\!^\mathrm{s}$429 &  +12$^\circ$18$^\prime$50\farcs43 & 6.0763$\pm$0.0005 &  \cii\   & --26.85 & 1, 2 \\
J1030+0524  & 10$^\mathrm{h}$30$^\mathrm{m}$27$.\!\!^\mathrm{s}$098 &  +05$^\circ$24$^\prime$55\farcs00 &  6.308$\pm$0.001  &  \mgii\  & --26.93 & 4, 2 \\
P159--02    & 10$^\mathrm{h}$36$^\mathrm{m}$54$.\!\!^\mathrm{s}$193 & --02$^\circ$32$^\prime$37\farcs85 & 6.3809$\pm$0.0005 &  \cii\   & --26.74 & 1, 2 \\
J1048--0109 & 10$^\mathrm{h}$48$^\mathrm{m}$19$.\!\!^\mathrm{s}$081 & --01$^\circ$09$^\prime$40\farcs45 & 6.6759$\pm$0.0005 &  \cii\   & --25.96 & 1, 5 \\
P167--13    & 11$^\mathrm{h}$10$^\mathrm{m}$33$.\!\!^\mathrm{s}$988 & --13$^\circ$29$^\prime$45\farcs84 & 6.5148$\pm$0.0005 &  \cii\   & --25.57 & 1, 6 \\
J1148+0702  & 11$^\mathrm{h}$48$^\mathrm{m}$03$.\!\!^\mathrm{s}$286 &  +07$^\circ$02$^\prime$08\farcs33 &  6.339$\pm$0.001  &  \mgii\  & --26.43 & 7, 2 \\
J1152+0055  & 11$^\mathrm{h}$52$^\mathrm{m}$21$.\!\!^\mathrm{s}$277 &  +00$^\circ$55$^\prime$36\farcs54 & 6.3643$\pm$0.0005 &  \cii\   & --25.08\tablenotemark{c} & 1, 2 \\
J1207+0630  & 12$^\mathrm{h}$07$^\mathrm{m}$37$.\!\!^\mathrm{s}$428 &  +06$^\circ$30$^\prime$10\farcs17 & 6.0366$\pm$0.0009 &  \cii\   & --26.57 & 1, 2 \\
P183+05     & 12$^\mathrm{h}$12$^\mathrm{m}$26$.\!\!^\mathrm{s}$974 &  +05$^\circ$05$^\prime$33\farcs59 & 6.4386$\pm$0.0004 &  \cii\   & --26.99 & 1, 6 \\
J1306+0356  & 13$^\mathrm{h}$06$^\mathrm{m}$08$.\!\!^\mathrm{s}$284 &  +03$^\circ$56$^\prime$26\farcs25 & 6.0337$\pm$0.0004 &  \cii\   & --26.76 & 1, 2 \\
P217--16    & 14$^\mathrm{h}$28$^\mathrm{m}$21$.\!\!^\mathrm{s}$371 & --16$^\circ$02$^\prime$43\farcs73 & 6.1498$\pm$0.0011 &  \cii\   & --26.89 & 1, 2 \\
J1509--1749 & 15$^\mathrm{h}$09$^\mathrm{m}$41$.\!\!^\mathrm{s}$781 & --17$^\circ$49$^\prime$26\farcs68 & 6.1225$\pm$0.0007 &  \cii\   & --27.09 & 1, 2 \\
P231--20    & 15$^\mathrm{h}$26$^\mathrm{m}$37$.\!\!^\mathrm{s}$841 & --20$^\circ$50$^\prime$00\farcs66 & 6.5864$\pm$0.0005 &  \cii\   & --27.14 & 1, 6 \\
P308--21    & 20$^\mathrm{h}$32$^\mathrm{m}$10$.\!\!^\mathrm{s}$003 & --21$^\circ$14$^\prime$02\farcs25 & 6.2341$\pm$0.0005 &  \cii\   & --26.30 & 1, 2 \\
J2100--1715 & 21$^\mathrm{h}$00$^\mathrm{m}$54$.\!\!^\mathrm{s}$707 & --17$^\circ$15$^\prime$21\farcs88 & 6.0812$\pm$0.0005 &  \cii\   & --25.50 & 1, 2 \\
J2211--3206 & 22$^\mathrm{h}$11$^\mathrm{m}$12$.\!\!^\mathrm{s}$417 & --32$^\circ$06$^\prime$12\farcs54 & 6.3394$\pm$0.0010 &  \cii\   & --26.65 & 1, 3 \\
P340--18    & 22$^\mathrm{h}$40$^\mathrm{m}$48$.\!\!^\mathrm{s}$978 & --18$^\circ$39$^\prime$43\farcs62 &   6.01$\pm$0.05   & Template & --26.36 & 2, 2 \\
J2318--3113 & 23$^\mathrm{h}$18$^\mathrm{m}$18$.\!\!^\mathrm{s}$393 & --31$^\circ$13$^\prime$46\farcs56 & 6.4435$\pm$0.0005 &  \cii\   & --26.06 & 1, 3 \\
J2318--3029 & 23$^\mathrm{h}$18$^\mathrm{m}$33$.\!\!^\mathrm{s}$099 & --30$^\circ$29$^\prime$33\farcs51 & 6.1458$\pm$0.0004 &  \cii\   & --26.16 & 1, 3 \\
P359--06    & 23$^\mathrm{h}$56$^\mathrm{m}$32$.\!\!^\mathrm{s}$439 & --06$^\circ$22$^\prime$59\farcs18 & 6.1722$\pm$0.0004 &  \cii\   & --26.74 & 1, 2 \\
\enddata
\tablenotetext{a}{Method used to determine the redshift, with ``Template" referring to a template fit to the rest-frame UV quasar spectrum.}
\tablenotetext{b}{References for the redshift and $M_{1450}$: (1) \citet{dec18}, (2) \citet{ban16}, (3) B.\,P.\ Venemans et al.\ 2018, in preparation, (4) \citet{kur07}, (5) \citet{fwan17}, (6) \citet{maz17b}, (7) \citet{jia16}.}
\tablenotetext{c}{After the creating the sample, additional analysis of the optical spectrum of this quasar decreased the absolute magnitude below our luminosity cut (Section~\ref{sec:sample}).} 
\end{deluxetable*}

To study the range of FIR properties displayed in the host galaxies of $z\sim6$ quasars, we created a luminosity-limited sample. Based on the luminosity limit of quasars found in wide area sky surveys \citep[e.g.,][]{ban16,jia16}, we selected quasars with an absolute UV magnitude brighter than $M_{1450} < -25.25$, approximately corresponding to a black hole mass $M_\mathrm{BH}>2.5\times10^8$\,\msun\ (assuming Eddington accretion). We further set a lower redshift cut of $z>5.94$ to ensure that the \cii\,158\,$\mu$m emission line is redshifted to the easily accessible 1.2\,mm band (ALMA band 6). Finally, to allow observations at low airmass with ALMA, we set a declination limit of $< +15^{\circ}$. In 2015 April (when we created the sample), the sample consisted of 43 quasars, of which 9 were unpublished at the time. 

Of the 43 quasars in our luminosity-limited sample, 8 were already observed in \cii\ with sensitive interferometers, such as ALMA and IRAM/PdBI. The final target list for our ALMA quasar survey thus consisted of 35 sources\footnote{Additional analysis of one of the quasars in our sample, J1152+0055, by \citet{ban16} resulted in an absolute magnitude 0.17\,mag below our limit of $M_{1450}=-25.25$ (Table~\ref{tab:sample}). 
We decided to keep it in our sample.}. Of these, 27 were observed with ALMA in Cycle 3 \citep[see][for more details]{dec18}. The remaining 8 sources were not observed, mostly due to poor visibility when ALMA was in a suitable array configuration. The coordinates, redshifts, and optical properties of the 27 observed quasars are listed in Table~\ref{tab:sample} and a brief description of the observations is given in the next section.

\subsection{ALMA Observations and Analysis}
\label{sec:observations}

\begin{deluxetable*}{lccccccccc}[th]
\tablecaption{Measured and Derived Properties of the Quasars in our Sample \label{tab:prop}}
\tablewidth{0pt}
\tablehead{\colhead{Name} & \colhead{S/N} & \colhead{S/N} &
\colhead{$S_\mathrm{rest,\,1790\,GHz}$} &
\colhead{$S_\mathrm{rest,\,1900\,GHz}$} &
\colhead{size} & \colhead{size} &
\colhead{log \lfir\tablenotemark{a}} & \colhead{SFR\tablenotemark{a,b}} & \colhead{log \mdust\tablenotemark{a}} \\
\colhead{} & \colhead{(at 1790\,GHz)} &
\colhead{(at 1900\,GHz)} & \colhead{(mJy)} &
\colhead{(mJy)} & \colhead{(at 1790\,GHz)} &
\colhead{(at 1900\,GHz)} & \colhead{(\lsun)} &
\colhead{(\msunyr)} & \colhead{(\msun)} }
\startdata
P007+04     & 28  & 26  & 3.280$\pm$0.220 & 3.880$\pm$0.260 &    0\farcs58$\times$0\farcs28 &    0\farcs54$\times$0\farcs23 & 12.87 & 1564 & 8.68 \\
P009--10     & 22  & 20  & 4.310$\pm$0.300 & 4.510$\pm$0.360 &    0\farcs86$\times$0\farcs53 &    0\farcs72$\times$0\farcs46 & 12.94 & 1808 & 8.74 \\
J0046--2837  & 1   & 2   & \multicolumn{2}{c}{0.128$\pm$0.039\tablenotemark{c} (3.3$\sigma$)} &           --            &           --           & 11.43 & 56 & 7.23 \\
J0142--3327  & 21  & 19  & 1.810$\pm$0.140 & 2.540$\pm$0.180 &    0\farcs92$\times$0\farcs36 &    0\farcs78$\times$0\farcs57 & 12.73 & 1130 & 8.54 \\
P065--26     & 13  & 14  & 1.040$\pm$0.130 & 1.650$\pm$0.140 &    0\farcs73$\times$0\farcs25 &    0\farcs69$\times$0\farcs37 & 12.53 & 702 & 8.33 \\
P065--19     & 7   & 8   & 0.561$\pm$0.124 & 0.482$\pm$0.094 &    0\farcs97$\times$0\farcs84 & $<$1\farcs08$\times$0\farcs71 & 11.98 & 202 & 7.79 \\
J0454--4448  & 12  & 12  & 0.672$\pm$0.080 & 0.992$\pm$0.137 &    0\farcs62$\times$0\farcs22 &    0\farcs62$\times$0\farcs48 & 12.30 & 417 & 8.10 \\
J0842+1218  & 10  & 6   & 0.542$\pm$0.064 & 0.732$\pm$0.223 &    0\farcs65$\times$0\farcs10 &    1\farcs29$\times$0\farcs15 & 12.16 & 304 & 7.97 \\
J1030+0524  & 2   & 2   & \multicolumn{2}{c}{0.134$\pm$0.046\tablenotemark{c} (2.9$\sigma$)} &           --            &           --           & 11.49 & 64 & 7.29 \\
P159--02     & 11  & 11  & 0.646$\pm$0.086 & 0.679$\pm$0.091 &    0\farcs73$\times$0\farcs44 & $<$1\farcs23$\times$0\farcs94 & 12.16 & 305 & 7.97 \\
J1048--0109  & 41  & 38  & 2.722$\pm$0.094 & 3.110$\pm$0.120 &    0\farcs61$\times$0\farcs33 &    0\farcs53$\times$0\farcs30 & 12.85 & 1500 & 8.66 \\
P167--13     & 11  & 14  & 0.749$\pm$0.091 & 1.071$\pm$0.092 &    1\farcs02$\times$0\farcs71 &    0\farcs95$\times$0\farcs46 & 12.38 & 502 & 8.18 \\
J1148+0702  & 4   & 7   & 0.494$\pm$0.168 & 0.664$\pm$0.181 &    1\farcs89$\times$0\farcs80 &    1\farcs37$\times$0\farcs30 & 12.15 & 293 & 7.95 \\
J1152+0055  & 2   & 4   & \multicolumn{2}{c}{0.124$\pm$0.043\tablenotemark{c} (2.9$\sigma$)} &           --            &           --           & 11.45 & 60 & 7.26 \\
J1207+0630  & 6   & 5   & 0.407$\pm$0.064 & 0.467$\pm$0.091 &           --            &           --           & 11.96 & 192 & 7.77 \\
P183+05     & 42  & 41  & 4.770$\pm$0.140 & 5.850$\pm$0.160 &    0\farcs58$\times$0\farcs48 &    0\farcs62$\times$0\farcs40 & 13.11 & 2693 & 8.91 \\
J1306+0356  & 15  & 12  & 1.250$\pm$0.100 & 1.480$\pm$0.220 &    1\farcs20$\times$0\farcs37 &    1\farcs05$\times$0\farcs45 & 12.46 & 605 & 8.26 \\
P217--16     & 5   & 4   & 0.350$\pm$0.070 & 0.421$\pm$0.100 &           --            &           --           & 11.93 & 178 & 7.73 \\
J1509--1749  & 23  & 20  & 1.365$\pm$0.089 & 1.760$\pm$0.110 & $<$1\farcs44$\times$0\farcs93 & $<$1\farcs35$\times$0\farcs87 & 12.55 & 742 & 8.35 \\
P231--20     & 36  & 84  & 3.920$\pm$0.450 & 4.210$\pm$0.360 &    0\farcs61$\times$0\farcs52 &    0\farcs31$\times$0\farcs25 & 12.99 & 2026 & 8.79 \\
P308--21     & 17  & 6   & 0.846$\pm$0.080 & 0.824$\pm$0.203 &    0\farcs83$\times$0\farcs58 &    1\farcs67$\times$0\farcs73 & 12.23 & 358 & 8.04 \\
J2100--1715  & 7   & 5   & 0.554$\pm$0.140 & 0.877$\pm$0.248 &    $<$0\farcs78$\times$0\farcs66 & 0\farcs83$\times$0\farcs33 & 12.24 & 366 & 8.05 \\
J2211--3206  & 11  & 9   & 0.689$\pm$0.075 & 0.733$\pm$0.121 &    0\farcs58$\times$0\farcs29 & $<$0\farcs87$\times$0\farcs69 & 12.19 & 326 & 8.00 \\
P340--18     & 3   & 4   & \multicolumn{2}{c}{0.174$\pm$0.046\tablenotemark{c} (3.8$\sigma$)} &           --            &           --           & 11.56 & 76 & 7.36 \\
J2318--3113  & 5   & 5   & 0.418$\pm$0.087 & 0.567$\pm$0.105 &           --            &           --           & 12.10 & 261 & 7.90 \\
J2318--3029  & 26  & 24  & 3.190$\pm$0.200 & 3.930$\pm$0.220 &    0\farcs71$\times$0\farcs38 &    0\farcs79$\times$0\farcs35 & 12.90 & 1653 & 8.70 \\
P359--06     & 8   & 7   & 0.982$\pm$0.170 & 1.020$\pm$0.210 & $<$1\farcs14$\times$0\farcs64 &    0\farcs79$\times$0\farcs22 & 12.32 & 436 & 8.12 \\
\enddata
\tablenotetext{a}{Quoted uncertainties are measurement errors, assuming the dust spectral energy distribution can be described by a modified blackbody with $T_\mathrm{dust}=47$\,K, $\beta=1.6$, and a negligible dust optical depth at $\nu_\mathrm{rest}=1790$\,GHz. The actual uncertainties are dominated by our assumptions on the shape of the dust spectral energy distribution (see Sections~\ref{sec:lfir}--\ref{sec:mdust}) for a detailed discussion).}
\tablenotetext{b}{Assuming the dust is heated only by star formation (see Section~\ref{sec:sfrs}).}
\tablenotetext{c}{These flux densities were measured in a map created by averaging all four spectral windows.}
\end{deluxetable*}

The new ALMA observations (program ID: 2015.1.01115.S, PI: F.\ Walter) consisted of 8\,min on-source integrations with two (partly overlapping) bandpasses of 1.875\,GHz covering the redshifted \cii\,158\,$\mu$m line and two bandpasses of 1.875\,GHz width each targeting the quasar continuum at slightly lower frequencies. The typical beam had a size of $\sim$1\farcs0 and the typical rms noise is 0.5\,mJy\,beam$^{-1}$ in 30\,\kms\ bins. Further details of the observations and the data reduction can be found in \citet{dec18}. 

We generated two continuum maps for each source. One was created by averaging the line-free channels in the two spectral windows targeting the \cii\ line \citep[see][for details]{dec18}. This map provides the measurements of the continuum flux density of the quasar host at a rest-frame frequency of 1900\,GHz, $S_\mathrm{rest,\,1900\,GHz}$. The second map was constructed by averaging all the channels in the spectral windows in the lower sideband, covering a frequency typically $\sim$16\,GHz lower than that of the \cii\ line. The flux densities measured in this map will be referred to as $S_\mathrm{rest,\,1790\,GHz}$. For objects that were not detected in these maps, we created a continuum map by averaging all line-free channels of the four bandpasses (see below). 

\begin{figure*}
\includegraphics[width=0.95\textwidth]{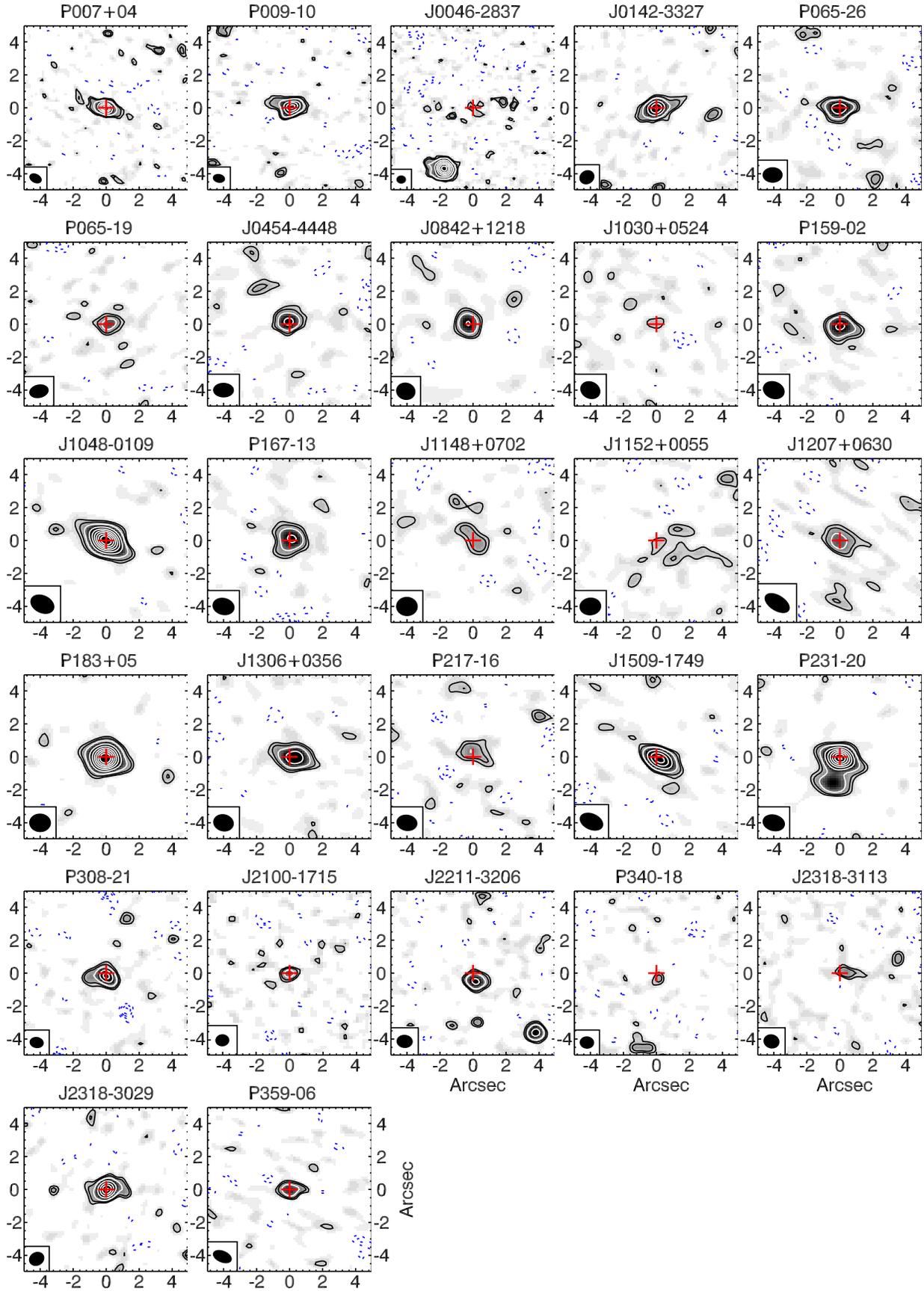}
\caption{Continuum maps of the 27 quasars observed in our survey. The postage stamps are 10\arcsec\,$\times$\,10\arcsec\ in size. The maps were created by averaging the channels in the spectral windows in the lower sideband, away from the \cii\ line,  probing a rest-frame frequency around 1790\,GHz. The optical positions of the quasars are indicated with a red cross. The dashed contours are $-$3$\sigma$ and $-$2$\sigma$ and the solid contours are +2$\sigma$, +3$\sigma$ and [5, 10, 15, 20, 25, 30, 35]\,$\times\sigma$.}
\label{fig:maps}
\end{figure*}

To determine the continuum flux densities of the quasar host galaxies, we performed the following steps. First, the peak flux density in the maps was measured by selecting the brightest pixel within 0\farcs5 of the quasar position (Table~\ref{tab:sample}). We also measured the source brightness and extent using the CASA task ``imfit." If the S/N ratio of the peak flux density in one of the continuum maps was at least 7, then the integrated flux densities provided by ``imfit'' were taken as the brightness of the source as listed in Table~\ref{tab:prop}. In the other cases a brightness equal to the peak flux density was adopted as the fits provided by ``imfit'' became poorly constrained. In four cases, for J0046--2837, J1030+0524, J1152+0055, and P340--18, the S/N of the peak pixel was below 3 in our deeper continuum map (the one covering the frequencies around 1790\,GHz in the rest-frame). In these cases, we created an additional continuum map by averaging all line-free channels in all four spectral windows and determined the peak flux density within 0\farcs5 of the quasar position. In these new maps, the four faint quasar hosts were potentially detected at a significance of $\sim$3$\sigma$--4$\sigma$ (Table~\ref{tab:prop}). In Figure~\ref{fig:maps} we show postage stamps for our sample.

\subsection{Literature Sample}
\label{sec:literature}

From the literature we collected all available observations of high-redshift ($z>5.7$) quasars obtained in the 1\,mm band. These observations include both bolometer observations \citep[with the IRAM 30m/MAMBO,][]{ber03a,pet03,wan07,wan08b,wan11b,omo13} and interferometric observations with ALMA \citep{wan13,wan16,wil13,wil15,wil17,ven16,ven17a} and the PdBI/NOEMA \citep{gal14,ban15b,maz17b,ven17c}. In total, 64 quasars with mm observations were found in the literature, of which 30 were detected. Most of the non-detections came from the bolometer observations, which typically had 1\,$\sigma$ noise levels of $0.5-1.0$\,mJy. A summary of the optical and FIR properties of the quasars with literature measurements is listed in Table~\ref{tab:lit}.

\section{Results}
\label{sec:results}

We detect all 27 quasar hosts in the dust continuum (Figure~\ref{fig:maps}) at a significance of $\gtrsim3$$\sigma$, of which 21 (78\%) have a peak flux density with an S/N\,$>5$. The flux densities span a range of a factor $\sim$50, from 0.12 to $\sim$6\,mJy. In several fields, additional sources adjacent to the quasars are visible. These objects are discussed in separate papers \citep[][]{dec17,cha18}. In Table~\ref{tab:prop} we list the continuum brightness of all the quasars and the measured sizes for the objects detected at S/N\,$>7$. In the next section (Section~\ref{sec:ext}) we will discuss the extent of the continuum emission. In Section~\ref{sec:dist} we will compare the properties of the quasars in our sample with those from the literature.

\subsection{Continuum emission size}
\label{sec:ext}

\begin{figure}
\includegraphics[width=\columnwidth]{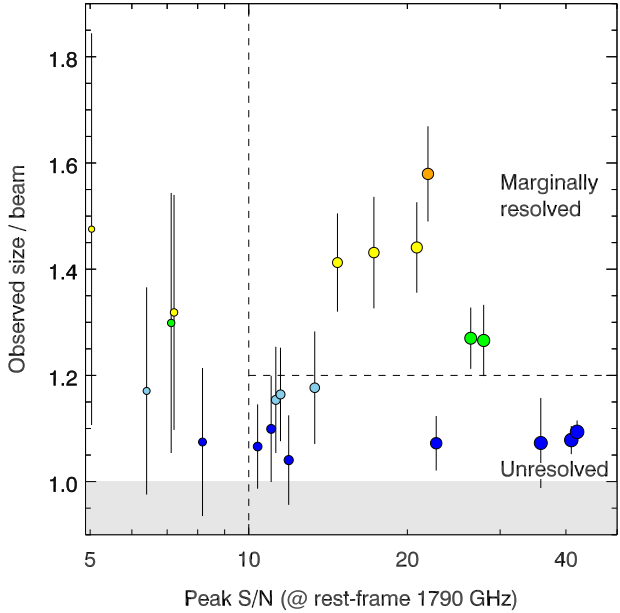}
\caption{Observed continuum size (major axis from Table~\ref{tab:prop}) divided by the beam size of the quasars in our sample as a function of peak S/N. The colors represent the ratio of observed size over the beam size. 
The size of the symbols scales with the S/N of the peak flux density. Sources with an observed size over the beam larger than $\sim$1.2 are considered marginally resolved.}
\label{fig:size}
\end{figure}

In Figure~\ref{fig:size} we show the ratio of the beam-convolved major axis of the continuum emission of the quasars and the beam size a function of the S/N of the continuum emission. Following \citet{dec18}, we only regard objects detected at S/N\,$>10$ as suitable to determine the extent of the emission. Of the quasars in our sample, 16 are detected with a S/N\,$>10$ at a rest-frame frequency of 1790\,GHz. At the resolution of our observations (typically 1\farcs1$\times$0\farcs9, or 6.2$\times$5.1\,kpc$^2$), none of the sources are resolved with measured sizes $>$2$\times$ the beam. Of the 16 high S/N sources, 10 (62.5\%) have an observed major axis less than 1.2 times the major axis of the beam. We consider these objects unresolved, as the measured size is within 1$\sigma$--2$\sigma$ of the size of the beam. Six quasars (37.5\%) are marginally resolved, with measured major axis size between 1.2 and 1.6 times that of the beam. The deconvolved sizes are in the range 3.3--6.9\,kpc, with significant uncertainties. To more accurately estimate the range of sizes of the quasar host galaxies, higher-resolution imaging is required \citep[e.g.,][]{sha17,ven17a}. Despite the large uncertainties, the sizes of the quasar host galaxies appear to be similar to those of star-forming galaxies at $z=2$ \citep[e.g.,][]{tad17}. Furthermore, we note that only a fraction ($\sim$15\%) of the sources studied here show signatures of a recent merger and/or a very nearby companion galaxy \citep[see][]{dec17}. This relatively small fraction seems to be at odds with the model that the AGN activity and the (obscured) star-formation are triggered by a merger \citep[e.g.,][]{san88,hop08,ale12}, although imaging with a higher spatial resolution is needed to exclude very close companions and/or very compact merger remnants.

\subsection{Sample properties}
\label{sec:dist}

\begin{figure}
\includegraphics[width=\columnwidth]{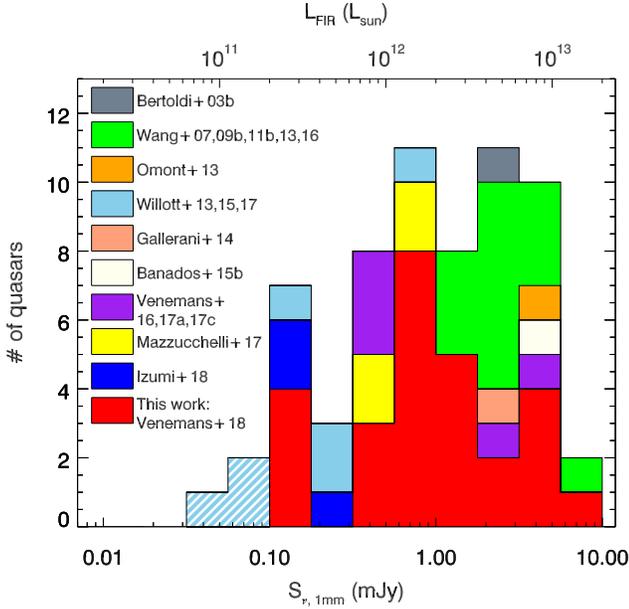}
\caption{Distribution of the observed brightness of the $z\gtrsim6$ quasar host galaxies in our sample (red histogram) at an observed wavelength (frequency) of $\sim$1\,mm ($\sim$250\,GHz). For illustration, we show the corresponding FIR luminosity on the top axis for a source at the median redshift of our sample ($z=6.1$) using a dust spectral energy distribution (SED) described by $T_d=47$\,K and $\beta=1.6$ (see Section~\ref{sec:lfir}). Also shown is the distribution of mm flux densities of quasars from the literature (Table~\ref{tab:lit}). Hashed bins indicate that the quasar host was not detected and the 3\,$\sigma$ limit is given. Bolometer upper limits (with a typical rms of $>$0.5\,mJy) are not shown.}
\label{fig:hist}
\end{figure}

In Figure~\ref{fig:hist} we show the histograms of observed continuum flux densities of the quasar host galaxies in our sample and from the literature. When compared to quasars previously observed at 1\,mm, our survey covers nearly the full range of flux densities.
The mean flux density of our sample is $\langle S_{\mathrm{obs,\,1\,mm}}\rangle = 1.6$\,mJy, very similar to the mean flux density of luminous $z\sim6$ quasars of 1.26\,mJy reported by \citet{wan08b} based on MAMBO bolometer observations. The mean flux density of our sample is dominated by a handful of bright quasars with $S_{\mathrm{obs,1\,mm}}>1$\,mJy. The median flux density of our sample is 0.88\,mJy, which is only slightly higher than the average flux density of $z\sim6$ quasar hosts that were not individually detected by MAMBO \citep[$S_\mathrm{obs,\,250\,GHz}=0.52\pm0.13$\,mJy,][]{wan08b}. This is not too surprising, as the \citet{wan08b} sample of quasars overlaps with ours (see Section~\ref{sec:sample}). 

A sample of, on average, less luminous (in the rest-frame UV) quasars from the CFHQS at $5.8<z<6.5$ was observed by \citet{omo13} using the MAMBO bolometer. Only a single source was detected at an S/N\,$>3$ and the stacked 1\.mm flux density of the sample was 
$\langle S_{\mathrm{obs,\,1\,mm}}\mathrm{(CFHQS)}\rangle = 0.41$\,mJy (after removing the single detection). Although the average 1\,mm flux density of the CFHQS quasars is lower than the average of our sample, the uncertainties are high. In Section~\ref{sec:corr} we will discuss possible correlations between the (UV) luminosity of the quasars and the brightness at millimeter wavelengths.

\section{Discussion}
\label{sec:discussion}

\subsection{The FIR Luminosity of $z>6$ Quasar Host Galaxies}
\label{sec:lfir}

The shape of the FIR continuum of high-redshift quasars is often assumed to follow a modified blackbody \citep[e.g.,][]{pri01,bee06,lei14}:

\begin{equation}
S_\nu \propto (1-e^{-\tau_\mathrm{dust}})~\frac{3h\nu^{3}/\,c^2}{e^{h\nu/(k T_\mathrm{dust})} - 1},
\label{eq:sdust}
\end{equation}

\noindent
with $\nu$ being the rest-frame frequency, $\tau_\mathrm{dust}=(\nu/\nu_0)^\beta$ being the dust optical depth, $\beta$ being the dust emissivity power-law spectral index, and $T_\mathrm{dust}$ being the dust temperature. The FIR luminosity, \lfir, is calculated by integrating Equation~(\ref{eq:sdust}) between 42.5 and 122.5\,$\mu$m in the rest-frame \citep[e.g.,][]{hel88}.

\begin{figure}
\includegraphics[width=\columnwidth]{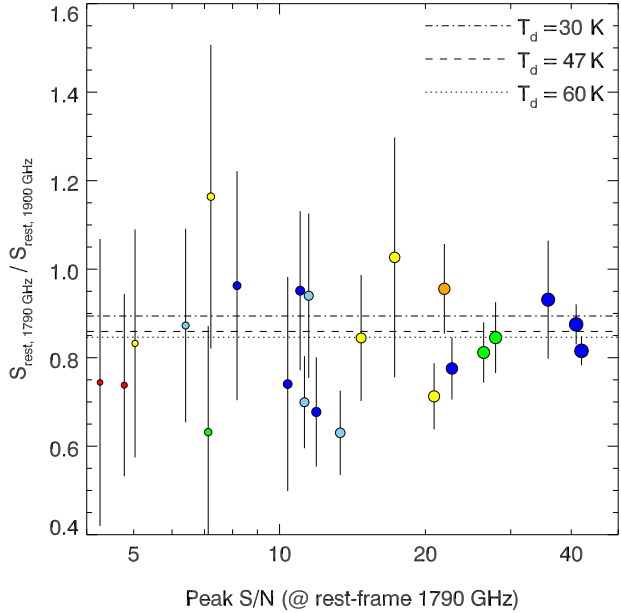}
\caption{Ratio of the flux density at a rest-frame frequency of $\sim$1790\,GHz to that at a rest-frame frequency of 1900\,GHz as a function of S/N. The size of the symbol scales with the S/N and the colors represent the size of the emission as plotted in Figure~\ref{fig:size}. The dot--dashed, dashed, and dotted lines indicate the ratio for a modified blackbody at $z=6.1$ (the median redshift of our sample) with an emissivity index of $\beta=1.6$ and a dust temperature of \tdust\,=30, 47, and 60\,K, respectively. Within 2$\sigma$, the measurements are consistent with the fiducial dust properties ($T_\mathrm{dust}=47$\,K and $\beta=1.6$).}
\label{fig:contratio}
\end{figure}

To compute the FIR luminosity we need to make assumptions on the dust temperature, spectral index, and dust optical depth. As described in Section~\ref{sec:observations} the ALMA observations provide measurements of the dust continuum at two different rest-frame frequencies, at roughly 1790 and 1900\,GHz. The ratio of these two measurements could constrain the dust spectral energy distribution \citep[see, for example, the discussion in][]{ven16}. In Figure~\ref{fig:contratio} we plot the ratio of the 1790 and 1900\,GHz flux densities as a function of S/N. As expected for the dust temperatures considered here (\tdust\,$>30$\,K, see below), the continuum at 1900\,GHz in the rest-frame is brighter than that at 1790\,GHz. Within 2$\sigma$, the $S_{\mathrm{rest},\,\mathrm{1790\,GHz}}/S_{\mathrm{rest},\,\mathrm{1900\,GHz}}$ of our quasar hosts with an S/N\,$>3$ at 1790\,GHz is $\sim$0.86, the expected ratio for a fiducial modified blackbody with \tdust\,$=47$\,K and $\beta=1.6$ \citep[the best-fit values by][see below]{bee06}. However, given that we are probing the continuum on the Rayleigh--Jeans tail of the dust emission, from our data alone we cannot derive a dust temperature together with the emissivity index, as they are degenerate. To accurately constrain the characteristics of the dust emission, we require continuum measurements at different frequencies. 

Following \citet{bee06} and \citet{ven16}, here we assume that the dust optical depth is negligible at far-infrared frequencies, i.e.\ $\tau_\mathrm{dust}\ll1$ at rest-frame frequencies $\nu_\mathrm{rest}<7.5$\,THz. It should be noted that an analysis of the dust SED of high-redshift submillimeter galaxies found that the dust optical depth can be significant at rest-frame frequencies $\nu_\mathrm{rest}<2$\,THz \citep[e.g.,][]{rie13,rie14,spi16}. This would modify the derived parameters for our quasar hosts. For example, if the dust optical depth is $\tau_\mathrm{dust}=1$ at $\nu_\mathrm{rest}=1790$\,GHz, then for a modified blackbody with \tdust\,$=35-55$\,K and $\beta=1.6$ (see below) the derived FIR luminosities will be lower by a factor of 2--3. 

In the literature, several groups have measured the (average) dust temperature and emissivity index of quasar hosts. \citet{pri01} found a dust temperature of \tdust\,$=41$\,K and an emissivity index of $\beta=1.95$ for a sample of $z\approx4$ quasars, while \citet{bee06} measured \tdust\,$=47$\,K and $\beta=1.6$. More recently, \citet{sta18} parameterized the dust spectral energy distribution of a sample of gravitationally lensed quasars at $z=1-4$ with \tdust\,$=38$\,K and $\beta=2.0$. 

Alternatively, we can fit templates of the local star-forming galaxies Arp 220 and M82 \citep[from e.g.,][]{sil98}, bypassing the various uncertainties introduced when using Equation~(\ref{eq:sdust}). To give an example, fitting the dust spectral energy distribution of Arp 220 with Equation~(\ref{eq:sdust}) results in a high dust temperature of \tdust\,$=66$\,K and a significant dust optical depth of $\tau_\mathrm{dust}\approx2$ at a rest-frame frequency of 1900\,GHz \citep[e.g.,][]{ran11}. 

Following the literature on $z\sim6$ quasars \citep[e.g.,][]{wan08b,wan13,wil13,ven16}, we here assume that the dust spectral energy distribution can be described by a modified blackbody with a dust temperature of \tdust\,$=47$\,K and an emissivity index of $\beta=1.6$ \citep[e.g.,][]{bee06,lei14}. In Table~\ref{tab:prop} we list the derived properties of the quasar hosts in our sample. As discussed above, the listed values in the table highly depend on the assumptions on the dust temperature and emissivity made here. For example, if we instead assume the best-fit values from \citet{pri01}, the FIR luminosity is lower by 11\%. Using \tdust\,$=38$\,K and $\beta=2.0$ results in an FIR luminosity that is 23\% lower, while with a higher dust temperature of \tdust\,$=55$\,K (with $\beta=1.6$), as found for a quasar host galaxy at $z=6.4$ \citep[e.g.,][]{bee06}, the \lfir\ is 50\% higher. Scaling the M82 and Arp 220 templates to our measured flux density results in an FIR luminosity that is 38\% and 48\% lower, respectively.

The quasars in our ALMA sample have FIR luminosities between \lfir\,$=2.7\times10^{11}$\,\lsun\ and \lfir\,$=1.3\times10^{13}$\,\lsun, with a median value of \lfir\,$=1.8\times10^{12}$\,\lsun\ (Table~\ref{tab:prop}). Of the 27 quasars observed, 20 (74\%) have FIR luminosities above $10^{12}$\,\lsun\ (the classical definition of a ULIRG). The remaining 7 (26\%) have $10^{11}<$\,\lfir\,$<10^{12}$\,\lsun. If we include sources from the literature that fulfill our sample selection criteria (Section~\ref{sec:sample}) and have a detection at 1\,mm (Table~\ref{tab:lit}), then we derive a similar median FIR luminosity of \lfir\,$=2.1\times10^{12}$\,\lsun\ and a ULIRG fraction of 81\%. 

We stress again that the FIR luminosities derived above strongly depend on the assumed shape of the dust spectral energy distribution. Additional photometry at other FIR frequencies is needed to better constrain the FIR luminosity for our quasar host galaxies.

\subsection{SFRs}
\label{sec:sfrs}

For high-redshift quasars, the dominant heating source of the dust that produces the infrared radiation appears to be stars \citep[e.g.,][but see, e.g., \citealt{sch15}]{lei14,bar15,ven17b}. \nocite{sch15} To estimate the SFR in the $z\gtrsim6$ quasar hosts, we can therefore apply the scaling relation found in the local universe between the total infrared (TIR) luminosity, \ltir, and the SFR: SFR\,$=1.48\times10^{-10} L_\mathrm{TIR}/L_\sun$ \citep[][]{mur11}. The total infrared luminosity can be obtained by integrating the dust spectral energy distribution between the rest-frame wavelengths of 3 and 1100\,$\mu$m \citep{ken12}. 

Computing TIR luminosities for our quasar hosts assuming a dust spectral energy distribution parameterized by \tdust\,$=47$\,K and $\beta=1.6$ (see Section~\ref{sec:lfir}), we derive SFRs of SFR\,$=50-2700$\,\msunyr\ (Table~\ref{tab:prop}). This assumes all the dust emission is heated by star-formation, so these values can be considered upper limits on the obscured SFRs in the quasar host galaxies. We caution that the derived SFRs strongly depend on our assumed dust properties and the uncertainties on the SFRs reported in Table~\ref{tab:prop} are up to a factor of $\sim$2--3 (see the discussion in Section~\ref{sec:lfir}). As an example, if we assume a dust temperature of \tdust\,$=55$\,K (instead of \tdust\,$=47$\,K) the resulting range of SFR in our quasar host galaxies is SFR\,$=90-4550$\,\msunyr, while fitting an Arp\,220 template to our measured flux densities (see Section~\ref{sec:lfir}) lead to lower derived SFRs of SFR\,$=30-1650$\,\msunyr. For a comparison between the SFRs in the host galaxies derived from the dust emission and from the \cii\ emission line, we refer to the discussion in \citet[][see their Figure 9]{dec18}. We caution that the derived SFRs strongly depend on our assumed dust properties and the uncertainties on the SFRs reported in Table~\ref{tab:prop} are up to a factor of $\sim$3 (see the discussion in Section~\ref{sec:lfir}).

Using the (highly uncertain) size estimates reported in Table~\ref{tab:prop}, we estimate SFR surface densities (SFRD) ranging from SFRD\,=\,$8-376$\,\msunyr\,kpc$^{-2}$. Given the low spatial resolution of our data, the measured sizes should be considered upper limits and the SFRD could be significantly higher. Higher spatial resolution observations will provide more accurate source sizes and better constrain the SFRD \citep[see, e.g.,][]{wal09b}.

\subsection{Dust Masses in Quasar Host at $z>6$}
\label{sec:mdust}

From the measured FIR flux density and assuming a dust temperature and emissivity index, we can estimate the total mass in dust, \mdust, using the equation:

\begin{equation}
M_\mathrm{dust} = \frac{S_\nu D_L^2}{(1+z) \kappa_\nu(\beta) B_\nu(\nu, T_\mathrm{dust})},
\label{eq:mdust}
\end{equation}

\noindent
where $D_L$ is the luminosity distance, $\kappa_\nu(\beta)$ is the dust mass opacity coefficient, and $B_\nu$ is the Planck function. The opacity coefficient is given by $\kappa_\nu(\beta)=0.77\,(\nu/352\,\mathrm{GHz})^\beta$\,cm$^2$\,g$^{-1}$ \citep[e.g.,][]{dun00}. 

Assuming $T_\mathrm{dust}=47$\,K, $\beta=1.6$ and a negligible dust optical depth at $\nu_\mathrm{rest}=1790$\,GHz (the canonical values used in the literature for $z\sim6$ quasar hosts, see Section~\ref{sec:lfir}), the estimated dust masses range from $M_\mathrm{dust} \approx 2\times10^7$\,\msun\ for the faintest sources to nearly $M_\mathrm{dust} \approx 10^9$\,\msun\ for the brightest quasar host galaxy (Table~\ref{tab:prop}). 

Similar to the derived FIR luminosity, these estimates of the dust mass in the quasar host galaxies are highly uncertain due to the unknown characteristics of the dust (see the discussion in Section~\ref{sec:lfir}). If the dust is parameterized by $T_\mathrm{dust}=41$\,K, $\beta=1.9$ ($T_\mathrm{dust}=38$\,K, $\beta=2.0$), instead of $T_\mathrm{dust}=47$\,K and $\beta=1.6$ as used above, then the derived dust masses in Table~\ref{tab:prop} are $\sim$14\% ($\sim$11\%) lower. On the other hand, scaling our measured flux densities to the templates of Arp 220 and M82 from \citet{sil98} results in derived dust masses that are 1.3--4.6$\times$ higher.

\subsection{Correlation between the FIR and UV Luminosity}
\label{sec:corr}

\begin{figure*}
\plottwo{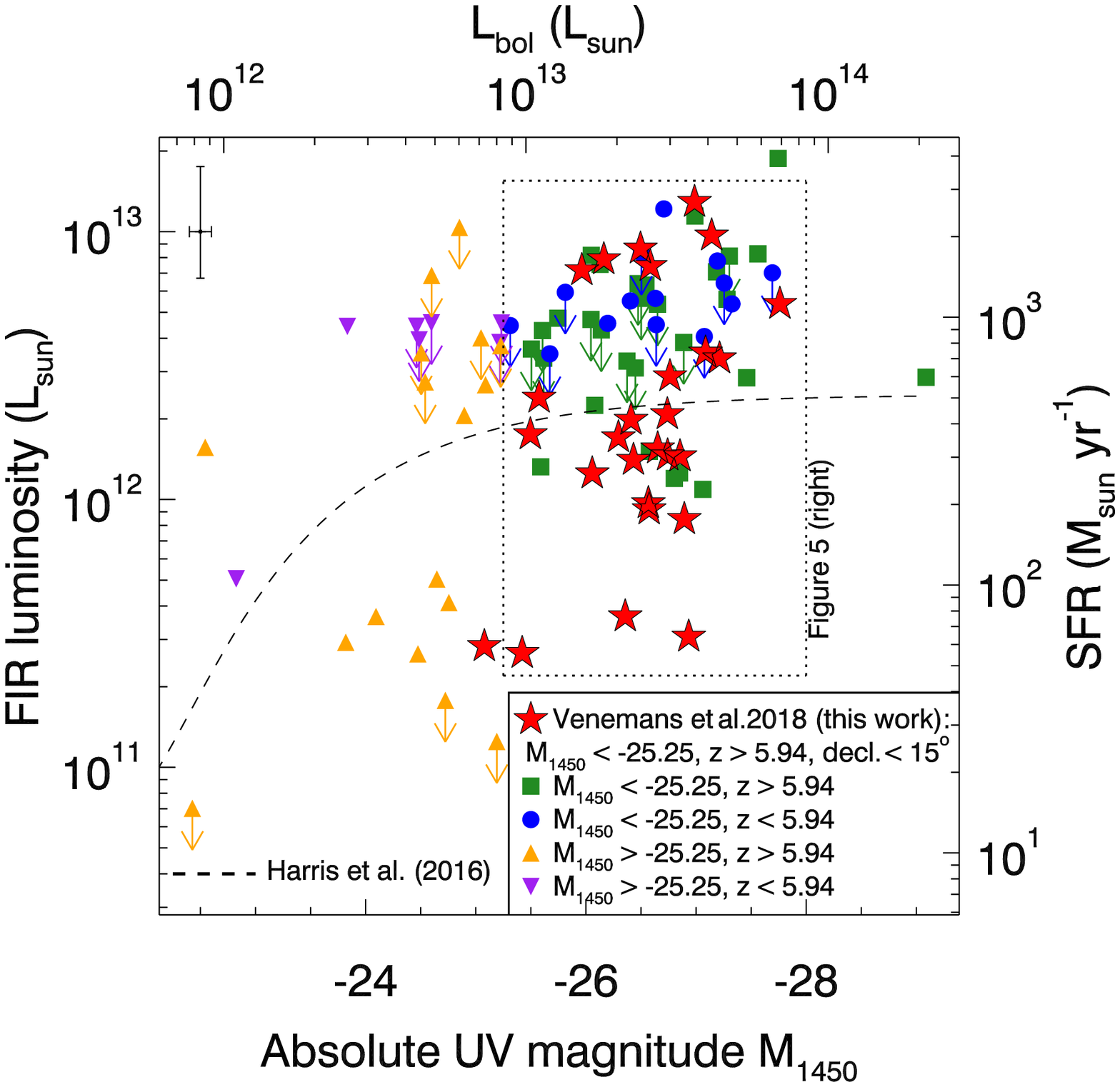}{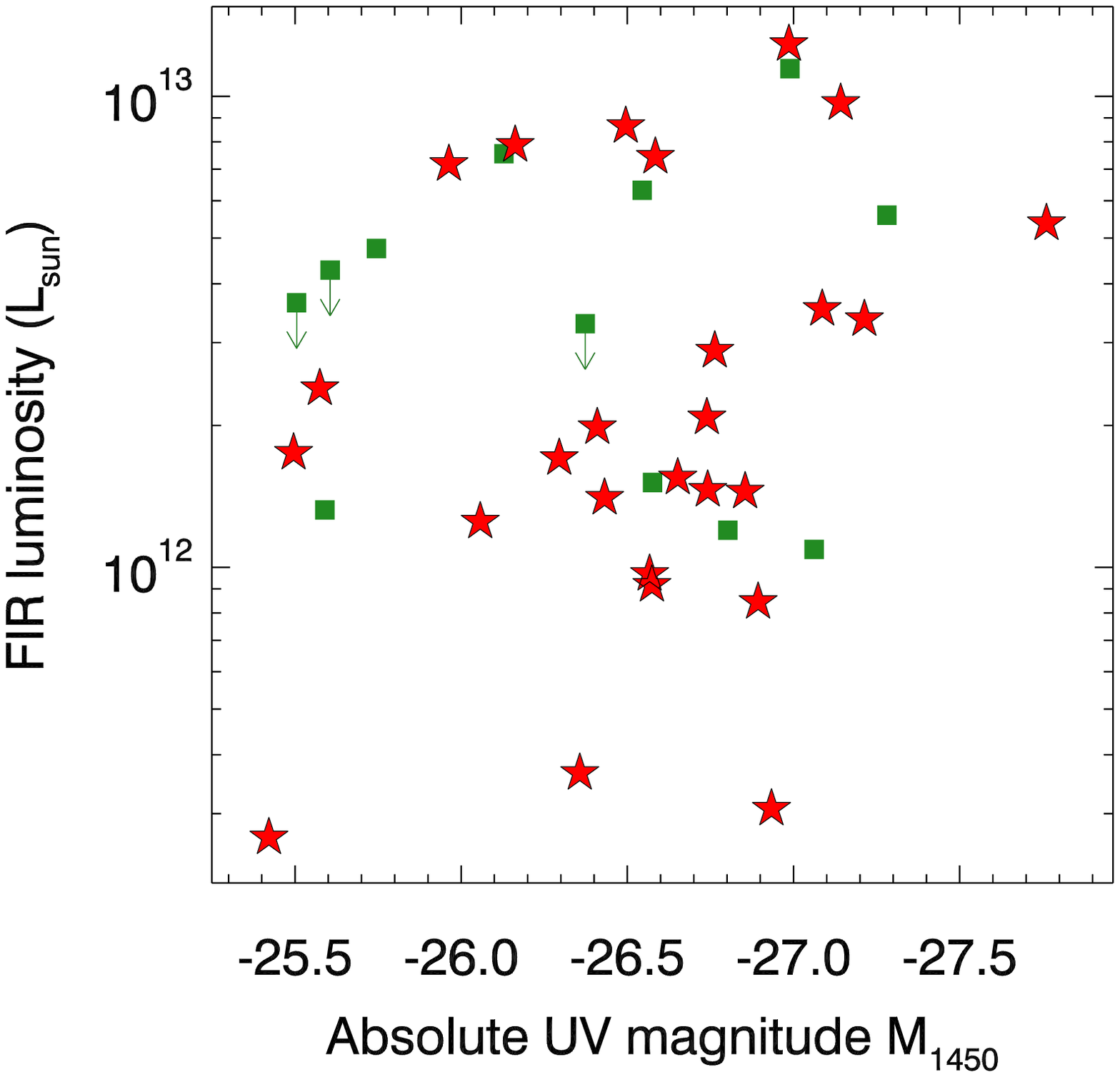}
\caption{Left: FIR luminosity (computed assuming \tdust\,=\,47\,K and $\beta=1.6$) of $z>5.7$ quasars as a function of the absolute magnitude $M_\mathrm{UV}$ at a wavelength 1450\,\AA\ in the rest-frame. Undetected objects are plotted with 3\,$\sigma$ upper limits (downward arrows). In the upper left corner the typical error bar is plotted. The dashed line is the relation between SFR and \lbol\ for quasars at $2<z<3$ \citep{har16}. A zoomed-in view of the dotted region is shown on the right. Right: same as the left plot, but this time for our quasar luminosity-limited ($M_{1450}<-25.25$) sample of $z\gtrsim6$ quasars. Within the small UV luminosity range probed by our survey, no correlation is evident between the brightness of the quasar and the luminosity of the dust emission in the host galaxy, with a large scatter in FIR luminosity for a given quasar brightness. }
\label{fig:contmuv}
\end{figure*}

In \citet{ven16} we collected all \cii\ and underlying dust continuum observations of $z\gtrsim6$ quasars from the literature and found that both the \cii\ emission and the FIR luminosity positively correlate with the luminosity of the AGN. However, the quasars used to derive that correlation were selected in different ways. With our homogenous, luminosity-limited quasar sample we can revisit this topic. In Figure~\ref{fig:contmuv} (left) we compare the FIR luminosity of our quasars (Table~\ref{tab:prop}) and all quasars from the literature (see Table~\ref{tab:lit}) with the brightness of the UV continuum emitted by the accreting black hole, $M_{1450}$ \citep[or, equivalently, the quasar bolometric luminosity \lbol, which is derived using Eq.~1 in][]{ven16}. Within the relatively narrow quasar luminosity range of our sample, the FIR and UV luminosities correlate only weakly, with a Pearson's $r$ of $r=-0.37$ \citep[a strong correlation is defined by us as $|r|>0.5$; see, e.g.,][]{ven16}. Quasars with the same UV luminosity can have FIR luminosities that differ by more than one order of magnitude. Similarly, at a given FIR luminosity, the range of $M_{1450}$ of the quasars is $>$3\,mag. Clearly, in the early universe there are quasars with rapid black hole growth but only little (obscured) stellar mass growth. At the same time, starburst galaxies with SFRs exceeding 1000\,\msunyr\ exist at $z>6$ that do not appear to have a highly accreting massive black hole \citep[e.g.,][]{rie13,mar18}. Fitting a straight line to the data results in the a correlation with a non-zero slope: log(\lfir)\,=\,$(12.41\pm0.05) - (0.18\pm0.04) \times (M_\mathrm{UV} + 26)$. There is a large dispersion around this fit, with a standard deviation of 0.45\,dex. 

Nonetheless, it is possible that this correlation is biased. For example, follow-up observations of the most UV luminous quasars were performed on the sources with bolometer detections. We therefore only plot the sources belonging to our luminosity-limited quasar sample ($M_\mathrm{1450}<-25.25$, $z>5.94$ and decl.\,$<15^\circ$, see Section~\ref{sec:sample}) in Figure~\ref{fig:contmuv} (right). In this complete sample, the correlation is even weaker, with a Pearson's $r=-0.22$. Similarly, in \citet{dec18}, we only find a weak correlation between $M_{1450}$ and the \cii\ luminosity, while the \cii-to-FIR luminosity ratio is independent of the quasar brightness. This argues against a strong contribution of the AGN to the heating of the dust, consistent with earlier conclusions based on different arguments \citep[e.g.,][]{lei14,dec18}.

These results are consistent with studies of quasars at lower redshifts. For example, \citet{har16} investigated the SFR in luminous quasars at $2<z<3$ and found that beyond a bolometric luminosity of \lbol$\gtrsim$$10^{13}$\,\lsun\ (corresponding roughly to $M_\mathrm{1450}\sim-25.5$) the SFR is independent of the brightness of the quasar (dashed line in Figure~\ref{fig:contmuv}). Similarly, \citet{pit16} found that the typical SFR remains constant for optically luminous quasars and does not vary with black hole mass or accretion rate.

\section{Summary}
\label{sec:summary}

In this paper we present ALMA snapshot observations (8\,minutes on-source) of 27 quasars at $z\gtrsim6$ selected from a UV luminosity-limited quasar sample. All quasars were detected in the dust continuum at an observed wavelength of $\sim$1\,mm, although the faintest quasars have only marginal detections (S/N\,$\approx3$). Below, we summarize our findings. The very high detection rate of our quasars \citep[100\% in the continuum and 85\% in \cii;][]{dec18} in very short, 8\,minutes, integration times will allow more detailed studies (e.g., multi-band SED, high spatial resolution observations) of these quasar host galaxies in the future. 

\begin{enumerate}

\item The quasar host galaxies in our survey span a wide range in observed millimeter continuum flux densities. The faintest quasar hosts have $S_\mathrm{obs,\,1\,mm}=0.12$\,mJy, which is among the faintest $z\sim6$ quasar hosts observed. The brightest quasar host in our survey, with $S_\mathrm{obs,\,1\,mm}=5.9$\,mJy, is the second most luminous quasar host after J2310+1855 at $z=6.0$ \citep{wan13}. The median flux density of quasar host galaxies in our survey is $S_\mathrm{obs,\,1\,mm}=0.9$\,mJy, very similar to the first bolometer results. 

\item As a result of the low spatial resolution of our observations (beam sizes around 1\arcsec, or $\sim$5.7\,kpc), 63\% of the quasar hosts detected at S/N\,$>10$ remain unresolved. The remaining quasar host are marginally resolved and have deconvolved sizes of 3.3--6.9\,kpc. 

\item The FIR luminosities, implied by the continuum measurements, are between \lfir\,$=3\times10^{11}$\,\lsun\ and \lfir\,$=1\times10^{13}$\,\lsun, assuming a dust temperature of 47\,K and an emissivity index of $\beta=1.6$. A high fraction of 70\% of quasars in our survey are hosted by ULIRGs. For a complete sample of quasars with $M_\mathrm{1450}<-25.25$ the fraction of ULIRGs is 78\%.

\item If the dust is heated by star-formation, the SFR implied by the infrared emission is 50--2700\,\msunyr. The derived dust masses in the quasar host galaxies are $M_\mathrm{dust}=2\times10^7-1\times10^9$\,\msun, implying significant amounts of dust and metals have been produced in these galaxies within 1\,Gyr after the Big Bang. 

\item Although the quasar hosts are marginally resolved at best, we can use the (upper limits on) the sizes to estimate star-formation rate densities (SFRDs). From the derived SFRs, we calculate SFRD\,$=10-400$\,\msunyr\,kpc$^{-2}$. These should be considered lower limits as the size of the continuum emission could be significantly smaller than the limits presented in Table~\ref{tab:prop}, see, e.g., \citet{wan13}, \citet{ven16}.

\end{enumerate}

In the local universe, a relation between the mass of a black hole and the bulge mass of the galaxy host has been reported \citep[e.g.,][]{kor13}. This has frequently been discussed in the context of a coevolution between the central black hole and the galaxy, i.e.\ that accretion onto the central black hole (black hole growth) should be accompanied by star-formation (stellar growth), but see \citet{jah11} for a different interpretation of this observational finding. 

The apparent lack of a correlation between black hole accretion (as measured by the UV luminosity) and stellar mass growth (as measured through the FIR luminosity) reported in this study could therefore indicate that black holes and the host galaxy do in fact not co-evolve, at least in the case of the most luminous quasars in the first Gyr of the universe.

This was also concluded by a different set of arguments by other studies of $z>6$ quasars, i.e.\ some dynamical mass estimates of quasar host galaxies indicate that they are under--massive compared to the supermassive black holes that they host \citep[e.g.,][but see, e.g., \citealt{wil15} for lower--luminosity counterexamples]{wal04,ven16,dec18}. 

There are several other interpretations of the apparent lack of correlation between black hole accretion and star-formation. For example, extinction of the UV quasar emission along the line of sight could introduce a significant amount of scatter. 
Similarly, if quasar host galaxies have a range of dust temperatures, our assumption that all hosts have $T_\mathrm{dust}=47$\,K will result in additional scatter. 
Alternatively, a more physical interpretation is that the timescales of black hole accretion could be much smaller than those of star-formation. The measured radiation due to accretion onto a black hole can in principle vary over a timescale of years \citep[given the small size of the emitting broad line region; see, e.g.,][]{kru17}, whereas the star-formation rate tracer used here, the FIR emission, has much longer timescales, $\gtrsim$$10^8$\,year \citep[e.g.,][]{hic12}. In this scenario, the weak correlation between SFR and quasar luminosity is probably mainly caused by the variability of the central source \citep[see, e.g.,][]{hic14}. Finally, the absence of a clear correlation could also be explained by an evolutionary scenario that links infrared luminous starburst galaxies and bright quasars \citep[e.g.,][]{san88,ale12}. In this scenario, the black hole grows rapidly in a dusty galaxy with a high SFR. The feedback of the luminous AGN removes the dusty interstellar medium, resulting in a UV-bright quasar. Simultaneously, the strong feedback removes the fuel for star-formation, and therefore suppresses the FIR luminosity \citep[e.g.,][]{lap14,man16}. With the available data it is difficult to differentiate between these different interpretations.

One way forward is to increase the UV luminosity range of the sample and investigate the host galaxy properties of fainter quasars. Studies of a small number of such low-luminosity quasars already suggest that the FIR luminosity of these quasars is on average lower \citep[e.g.,][]{wil13,wil17,izu18}, although, at the same time, there are galaxies with \lfir\,$>10^{13}$\,\lsun\ that do not show any black hole accretion \citep[e.g.,][]{rie13,mar18}. A better method to determine if the black hole and stellar mass of distant quasars grow simultaneously is to directly compare the mass of the black hole with the mass of stars in the host galaxy. 
In the near future, observations with the {\em James Webb Space Telescope} opens up the potential to detect the stars in the host galaxies. From the derived stellar masses it will be more straightforward to identify whether the first supermassive black holes already follow the local scaling relations.

\acknowledgments 
We thank the referee for providing valuable comments and suggestions. B.P.V., F.W., and E.P.F.\ acknowledge funding through the ERC grants ``Cosmic Dawn'' and ``Cosmic Gas." 
Part of the support for R.D.\ was provided by the DFG priority program 1573 ``The physics of the interstellar medium." 
D.R.\ acknowledges support from the National Science Foundation under grant number AST-1614213. 
This paper makes use of the following ALMA data:
ADS/JAO.ALMA\#2015.1.01115.S. ALMA is a partnership of ESO
(representing its member states), NSF (USA) and NINS (Japan), together
with NRC (Canada) and NSC and ASIAA (Taiwan), in cooperation with the
Republic of Chile. The Joint ALMA Observatory is operated by ESO,
AUI/NRAO, and NAOJ.

\vspace{5mm}
\facilities{ALMA}

\clearpage
\startlongtable
\begin{deluxetable*}{lcccccc}
\tablecaption{Quasars in the Literature with Observations at 1\,mm \label{tab:lit}}
\tablewidth{0pt}
\tablehead{ \colhead{Name} & \colhead{R.A.\ (J2000)} &
\colhead{Decl.\ (J2000)} & \colhead{Redshift} &
\colhead{$M_{1450}$} & \colhead{$S_\mathrm{obs,\,1\,mm}$ (mJy)} &
\colhead{References\tablenotemark{a}} }
\startdata
J0002+2550  & 00$^\mathrm{h}$02$^\mathrm{m}$39$.\!\!^\mathrm{s}$39  &  +25$^\circ$50$^\prime$34\farcs8  & 5.82~~ & --27.26 &     $<$2.64   & 1, 2, 3 \\
J0005--0006 & 00$^\mathrm{h}$05$^\mathrm{m}$52$.\!\!^\mathrm{s}$34  & --00$^\circ$06$^\prime$55\farcs8  & 5.844~ & --25.67 &     $<$1.44   & 4, 2, 5 \\
P006+39     & 00$^\mathrm{h}$24$^\mathrm{m}$29$.\!\!^\mathrm{s}$772 &  +39$^\circ$13$^\prime$18\farcs98 & 6.6210 & --26.85 & 0.55$\pm$0.18 & 6, 6, 6 \\
J0033--0125 & 00$^\mathrm{h}$33$^\mathrm{m}$11$.\!\!^\mathrm{s}$40  & --01$^\circ$25$^\prime$24\farcs9  & 6.13~~ & --25.09 & 1.13$\pm$0.36 & 7, 2, 5 \\
J0050+3445  & 00$^\mathrm{h}$50$^\mathrm{m}$06$.\!\!^\mathrm{s}$67  &  +34$^\circ$45$^\prime$22\farcs6  & 6.253~ & --26.65 &     $<$2.28   & 8, 2, 9 \\
J0055+0146  & 00$^\mathrm{h}$55$^\mathrm{m}$02$.\!\!^\mathrm{s}$92  &  +01$^\circ$46$^\prime$17\farcs8  & 6.0060 & --24.76 & 0.21$\pm$0.03 & 10, 2, 10 \\
J0100+2802  & 01$^\mathrm{h}$00$^\mathrm{m}$13$.\!\!^\mathrm{s}$02  &  +28$^\circ$02$^\prime$25\farcs8  & 6.3258 & --29.09 & 1.35$\pm$0.25 & 11, 2, 11 \\
J0102--0218 & 01$^\mathrm{h}$02$^\mathrm{m}$50$.\!\!^\mathrm{s}$64  & --02$^\circ$18$^\prime$09\farcs9  & 5.95~~ & --24.54 &     $<$1.14   & 12, 2, 9 \\
J0109--3047 & 01$^\mathrm{h}$09$^\mathrm{m}$53$.\!\!^\mathrm{s}$13  & --30$^\circ$47$^\prime$26\farcs32 & 6.7909 & --25.59 & 0.56$\pm$0.11 & 13, 2, 13 \\
J0129--0035 & 01$^\mathrm{h}$29$^\mathrm{m}$58$.\!\!^\mathrm{s}$51  & --00$^\circ$35$^\prime$39\farcs7  & 5.7787 & --23.83 & 2.57$\pm$0.06 & 14, 2, 14 \\
J0136+0226  & 01$^\mathrm{h}$36$^\mathrm{m}$03$.\!\!^\mathrm{s}$17  &  +02$^\circ$26$^\prime$05\farcs7  & 6.21~~ & --24.60 &     $<$2.91   & 15, 2, 9 \\
J0203+0012  & 02$^\mathrm{h}$03$^\mathrm{m}$32$.\!\!^\mathrm{s}$39  &  +00$^\circ$12$^\prime$29\farcs3  & 5.72~~ & --26.20 & 1.85$\pm$0.46 & 16, 2, 17 \\
J0210--0456 & 02$^\mathrm{h}$10$^\mathrm{m}$13$.\!\!^\mathrm{s}$19  & --04$^\circ$56$^\prime$20\farcs9  & 6.4323 & --24.47 & 0.12$\pm$0.04 & 18, 2, 18 \\
J0216--0455 & 02$^\mathrm{h}$16$^\mathrm{m}$27$.\!\!^\mathrm{s}$81  & --04$^\circ$55$^\prime$34\farcs1  & 6.01~~ & --22.43 &     $<$0.04   & 12, 2, 19 \\
J0221--0802 & 02$^\mathrm{h}$21$^\mathrm{m}$22$.\!\!^\mathrm{s}$718 & --08$^\circ$02$^\prime$51\farcs62 & 6.161~ & --24.65 & 0.25$\pm$0.05 & 8, 2, 19 \\
P036+03     & 02$^\mathrm{h}$26$^\mathrm{m}$01$.\!\!^\mathrm{s}$876 &  +03$^\circ$02$^\prime$59\farcs39 & 6.5412 & --27.28 & 2.50$\pm$0.50 & 20, 6, 20 \\
J0227--0605 & 02$^\mathrm{h}$27$^\mathrm{m}$43$.\!\!^\mathrm{s}$29  & --06$^\circ$05$^\prime$30\farcs2  & 6.20~~ & --25.23 &     $<$1.59   & 12, 2, 9 \\
J0239--0045 & 02$^\mathrm{h}$39$^\mathrm{m}$30$.\!\!^\mathrm{s}$24  & --00$^\circ$45$^\prime$05\farcs4  & 5.82~~ & --24.49 &     $<$1.62   & 21, 2, 17 \\
J0303--0019 & 03$^\mathrm{h}$03$^\mathrm{m}$31$.\!\!^\mathrm{s}$40  & --00$^\circ$19$^\prime$12\farcs9  & 6.078~ & --25.50 &     $<$1.53   & 22, 2, 5 \\
J0305--3150 & 03$^\mathrm{h}$05$^\mathrm{m}$16$.\!\!^\mathrm{s}$91  & --31$^\circ$50$^\prime$55\farcs94 & 6.6145 & --26.13 & 3.29$\pm$0.10 & 13, 2, 13 \\
J0316--1340 & 03$^\mathrm{h}$16$^\mathrm{m}$49$.\!\!^\mathrm{s}$87  & --13$^\circ$40$^\prime$32\farcs2  & 5.99~~ & --24.85 &     $<$4.32   & 15, 2, 9 \\
J0353+0104  & 03$^\mathrm{h}$53$^\mathrm{m}$49$.\!\!^\mathrm{s}$76  &  +01$^\circ$04$^\prime$05\farcs4  & 6.072~ & --26.37 &     $<$1.38   & 4, 2, 5 \\
J0818+1722  & 08$^\mathrm{h}$18$^\mathrm{m}$27$.\!\!^\mathrm{s}$40  &  +17$^\circ$22$^\prime$51\farcs8  & 6.02~~ & --27.46 & 1.19$\pm$0.38 & 1, 2, 5 \\
J0836+0054  & 08$^\mathrm{h}$36$^\mathrm{m}$43$.\!\!^\mathrm{s}$85  &  +00$^\circ$54$^\prime$53\farcs3  & 5.810~ & --27.69 &     $<$2.88   & 23, 2, 24 \\
J0840+5624  & 08$^\mathrm{h}$40$^\mathrm{m}$35$.\!\!^\mathrm{s}$09  &  +56$^\circ$24$^\prime$19\farcs9  & 5.85~~ & --27.19 & 3.20$\pm$0.64 & 25, 2, 3 \\
J0841+2905  & 08$^\mathrm{h}$41$^\mathrm{m}$19$.\!\!^\mathrm{s}$52  &  +29$^\circ$05$^\prime$04\farcs5  & 5.98~~ & --26.45 &     $<$1.29   & 1, 2, 5 \\
J0859+0022  & 08$^\mathrm{h}$59$^\mathrm{m}$07$.\!\!^\mathrm{s}$19  &  +00$^\circ$22$^\prime$55\farcs9  & 6.3903 & --24.09 & 0.16$\pm$0.02 & 26, 27, 26 \\
J0927+2001  & 09$^\mathrm{h}$27$^\mathrm{m}$21$.\!\!^\mathrm{s}$82  &  +20$^\circ$01$^\prime$23\farcs7  & 5.79~~ & --26.71 & 4.98$\pm$0.75 & 25, 2, 3 \\
J1044--0125 & 10$^\mathrm{h}$44$^\mathrm{m}$33$.\!\!^\mathrm{s}$04  & --01$^\circ$25$^\prime$02\farcs2  & 5.7847 & --27.33 & 3.12$\pm$0.09 & 14, 2, 14 \\
J1048+4637  & 10$^\mathrm{h}$48$^\mathrm{m}$45$.\!\!^\mathrm{s}$05  &  +46$^\circ$37$^\prime$18\farcs3  & 6.198~ & --27.19 & 3.00$\pm$0.40 & 4, 2, 28 \\
J1059--0906 & 10$^\mathrm{h}$59$^\mathrm{m}$28$.\!\!^\mathrm{s}$61  & --09$^\circ$06$^\prime$20\farcs4  & 5.92~~ & --25.81 &     $<$2.46   & 15, 2, 9 \\
J1120+0641  & 11$^\mathrm{h}$20$^\mathrm{m}$01$.\!\!^\mathrm{s}$465 &  +06$^\circ$41$^\prime$23\farcs81 & 7.0851 & --26.58 & 0.53$\pm$0.04 & 29, 2, 29 \\
J1137+3549  & 11$^\mathrm{h}$37$^\mathrm{m}$17$.\!\!^\mathrm{s}$73  &  +35$^\circ$49$^\prime$56\farcs9  & 6.03~~ & --27.30 &     $<$3.39   & 1, 2, 3 \\
J1148+5251  & 11$^\mathrm{h}$48$^\mathrm{m}$16$.\!\!^\mathrm{s}$64  &  +52$^\circ$51$^\prime$50\farcs2  & 6.4190 & --27.56 & 4.00$\pm$0.10 & 30, 2, 31 \\
J1202--0057 & 12$^\mathrm{h}$02$^\mathrm{m}$46$.\!\!^\mathrm{s}$37  & --00$^\circ$57$^\prime$01\farcs7  & 5.9289 & --22.83 & 0.25$\pm$0.01 & 26, 27, 26 \\
J1205--0000 & 12$^\mathrm{h}$05$^\mathrm{m}$05$.\!\!^\mathrm{s}$098 & --00$^\circ$00$^\prime$27\farcs97 & 6.730~ & --24.90 & 0.83$\pm$0.18 & 6, 6, 6 \\
J1250+3130  & 12$^\mathrm{h}$50$^\mathrm{m}$51$.\!\!^\mathrm{s}$93  &  +31$^\circ$30$^\prime$21\farcs9  & 6.15~~ & --26.47 &     $<$2.70   & 1, 2, 3 \\
J1319+0950  & 13$^\mathrm{h}$19$^\mathrm{m}$11$.\!\!^\mathrm{s}$29  &  +09$^\circ$50$^\prime$51\farcs4  & 6.1330 & --26.99 & 5.23$\pm$0.10 & 14, 2, 14 \\
J1335+3533  & 13$^\mathrm{h}$35$^\mathrm{m}$50$.\!\!^\mathrm{s}$81  &  +35$^\circ$33$^\prime$15\farcs8  & 5.95~~ & --26.63 & 2.34$\pm$0.50 & 25, 2, 3 \\
J1342+0928  & 13$^\mathrm{h}$42$^\mathrm{m}$08$.\!\!^\mathrm{s}$097 &  +09$^\circ$28$^\prime$38\farcs28 & 7.5413 & --26.80 & 0.41$\pm$0.07 & 32, 33, 32 \\
J1411+1217  & 14$^\mathrm{h}$11$^\mathrm{m}$11$.\!\!^\mathrm{s}$29  &  +12$^\circ$17$^\prime$37\farcs4  & 5.904~ & --26.64 &     $<$1.86   & 23, 2, 3 \\
J1425+3254  & 14$^\mathrm{h}$25$^\mathrm{m}$16$.\!\!^\mathrm{s}$30  &  +32$^\circ$54$^\prime$09\farcs0  & 5.85~~ & --26.40 & 2.27$\pm$0.51 & 34, 2, 5 \\
J1427+3312  & 14$^\mathrm{h}$27$^\mathrm{m}$38$.\!\!^\mathrm{s}$59  &  +33$^\circ$12$^\prime$42\farcs0  & 6.12~~ & --26.05 &     $<$1.98   & 35, 2, 5 \\
J1429+5447  & 14$^\mathrm{h}$29$^\mathrm{m}$52$.\!\!^\mathrm{s}$17  &  +54$^\circ$47$^\prime$17\farcs6  & 6.21~~ & --26.05 & 3.46$\pm$0.52 & 15, 2, 9 \\
J1436+5007  & 14$^\mathrm{h}$36$^\mathrm{m}$11$.\!\!^\mathrm{s}$74  &  +50$^\circ$07$^\prime$06\farcs9  & 5.85~~ & --26.50 &     $<$3.42   & 1, 2, 3 \\
J1602+4228  & 16$^\mathrm{h}$02$^\mathrm{m}$53$.\!\!^\mathrm{s}$98  &  +42$^\circ$28$^\prime$24\farcs9  & 6.09~~ & --26.89 &     $<$1.62   & 1, 2, 5 \\
J1621+5155  & 16$^\mathrm{h}$21$^\mathrm{m}$00$.\!\!^\mathrm{s}$70  &  +51$^\circ$55$^\prime$44\farcs8  & 5.71~~ & --27.07 &     $<$1.65   & 36, 2, 5 \\
J1623+3112  & 16$^\mathrm{h}$23$^\mathrm{m}$31$.\!\!^\mathrm{s}$81  &  +31$^\circ$12$^\prime$00\farcs5  & 6.2605 & --26.50 &     $<$2.40   & 37, 2, 3 \\
J1630+4012  & 16$^\mathrm{h}$30$^\mathrm{m}$33$.\!\!^\mathrm{s}$90  &  +40$^\circ$12$^\prime$09\farcs6  & 6.058~ & --26.14 &     $<$1.80   & 4, 2, 28 \\
J1641+3755  & 16$^\mathrm{h}$41$^\mathrm{m}$21$.\!\!^\mathrm{s}$64  &  +37$^\circ$55$^\prime$20\farcs5  & 6.047~ & --25.62 &     $<$1.41   & 8, 2, 9 \\
J2053+0047  & 20$^\mathrm{h}$53$^\mathrm{m}$21$.\!\!^\mathrm{s}$77  &  +00$^\circ$47$^\prime$06\farcs8  & 5.92~~ & --25.23 &     $<$1.89   & 21, 2, 17 \\
J2054--0005 & 20$^\mathrm{h}$54$^\mathrm{m}$06$.\!\!^\mathrm{s}$42  & --00$^\circ$05$^\prime$14\farcs8  & 6.0391 & --26.54 & 2.98$\pm$0.05 & 14, 2, 14 \\
P323+12     & 21$^\mathrm{h}$32$^\mathrm{m}$33$.\!\!^\mathrm{s}$191 &  +12$^\circ$17$^\prime$55\farcs26 & 6.5881 & --27.06 & 0.47$\pm$0.15 & 6, 6, 6 \\
J2147+0107  & 21$^\mathrm{h}$47$^\mathrm{m}$55$.\!\!^\mathrm{s}$40  &  +01$^\circ$07$^\prime$55\farcs0  & 5.81~~ & --25.31 &     $<$1.83   & 21, 2, 17 \\
J2216--0016 & 22$^\mathrm{h}$16$^\mathrm{m}$44$.\!\!^\mathrm{s}$47  & --00$^\circ$16$^\prime$50\farcs1  & 6.0962 & --23.82 & 0.14$\pm$0.03 & 26, 27, 26 \\
VIMOS2911   & 22$^\mathrm{h}$19$^\mathrm{m}$17$.\!\!^\mathrm{s}$227 &  +01$^\circ$02$^\prime$48\farcs88 & 6.1492 & --22.54 & 0.77$\pm$0.05 & 19, 2, 19 \\
J2229+1457  & 22$^\mathrm{h}$29$^\mathrm{m}$01$.\!\!^\mathrm{s}$66  &  +14$^\circ$57$^\prime$08\farcs30 & 6.1517 & --24.72 &     $<$0.09   & 10, 2, 10 \\
P338+29     & 22$^\mathrm{h}$32$^\mathrm{m}$55$.\!\!^\mathrm{s}$150 &  +29$^\circ$30$^\prime$32\farcs23 & 6.6660 & --26.08 & 0.97$\pm$0.22 & 6, 6, 6 \\
J2242+0334  & 22$^\mathrm{h}$42$^\mathrm{m}$37$.\!\!^\mathrm{s}$55  &  +03$^\circ$34$^\prime$21\farcs6  & 5.88~~ & --24.46 &     $<$1.83   & 15, 2, 9 \\
J2307+0031  & 23$^\mathrm{h}$07$^\mathrm{m}$35$.\!\!^\mathrm{s}$40  &  +00$^\circ$31$^\prime$49\farcs0  & 5.87~~ & --25.22 &     $<$1.59   & 21, 2, 17 \\
J2310+1855  & 23$^\mathrm{h}$10$^\mathrm{m}$38$.\!\!^\mathrm{s}$88  &  +18$^\circ$55$^\prime$19\farcs7  & 6.0031 & --27.75 & 8.91$\pm$0.08 & 14, 2, 14 \\
J2315--0023 & 23$^\mathrm{h}$15$^\mathrm{m}$46$.\!\!^\mathrm{s}$36  & --00$^\circ$23$^\prime$57\farcs5  & 6.12~~ & --25.61 &     $<$1.80   & 38, 2, 5 \\
J2318--0246 & 23$^\mathrm{h}$18$^\mathrm{m}$02$.\!\!^\mathrm{s}$80  & --02$^\circ$46$^\prime$34\farcs0  & 6.05~~ & --25.05 &     $<$1.68   & 12, 2, 9 \\
J2329--0301 & 23$^\mathrm{h}$29$^\mathrm{m}$08$.\!\!^\mathrm{s}$28  & --03$^\circ$01$^\prime$58\farcs8  & 6.4164 & --25.19 &     $<$0.06   & 19, 2, 19 \\
J2329--0403 & 23$^\mathrm{h}$29$^\mathrm{m}$14$.\!\!^\mathrm{s}$46  & --04$^\circ$03$^\prime$24\farcs1  & 5.90~~ & --24.60 &     $<$1.89   & 12, 2, 9 \\
J2348--3054 & 23$^\mathrm{h}$48$^\mathrm{m}$33$.\!\!^\mathrm{s}$35  & --30$^\circ$54$^\prime$10\farcs28 & 6.9018 & --25.75 & 1.92$\pm$0.14 & 13, 2, 13 \\
J2356+0023  & 23$^\mathrm{h}$56$^\mathrm{m}$51$.\!\!^\mathrm{s}$58  &  +00$^\circ$23$^\prime$33\farcs3  & 6.00~~ & --24.50 &     $<$1.47   & 21, 2, 17 \\
\enddata
\tablenotetext{a}{References for the redshift, $M_{1450}$, and $S_\mathrm{obs,\,1\,mm}$ data: (1) \citet{car10}, (2) \citet{ban16}, (3) \citet{wan07}, (4) \citet{der11}, (5) \citet{wan08b}, (6) \citet{maz17b}, (7) \citet{wil07}, (8) \citet{wil10b}, (9) \citet{omo13}, (10) \citet{wil15}, (11) \citet{wan16}, (12) \citet{wil09}, (13) \citet{ven16}, (14) \citet{wan13}, (15) \citet{wil10a}, (16) \citet{mor09}, (17) \citet{wan11b}, (18) \citet{wil13}, (19) \citet{wil17}, (20) \citet{ban15b}, (21) \citet{jia09}, (22) \citet{kur09}, (23) \citet{kur07}, (24) \citet{pet03}, (25) \citet{fan06b}, (26) \citet{izu18}, (27) \citet{mat18}, (28) \citet{ber03a}, (29) \citet{ven17a}, (30) \citet{wal09b}, (31) \citet{gal14}, (32) \citet{ven17c}, (33) \citet{ban18a}, (34) \citet{coo06}, (35) \citet{mcg06}, (36) \citet{jia16}, (37) \citet{wan11a}, (38) \citet{jia08}.}
\end{deluxetable*}

\end{document}